\documentclass[aps,pra,twocolumn,groupedaddress,amsmath,superscriptaddress]{revtex4}

\usepackage{amsfonts}
\usepackage{amssymb}
\usepackage{amsmath}
\usepackage[dvips]{graphicx}
\usepackage{color}

\usepackage{amsmath}
\usepackage{graphicx}
\usepackage{amsfonts}
\usepackage{amssymb}
\usepackage{hyperref}
\usepackage{color}
\usepackage{mathrsfs}
\usepackage{isomath}
\usepackage{amsmath}
\usepackage{amsthm}
\usepackage{epstopdf}
\usepackage{txfonts}

\begin{document}

\title{\hspace*{-.6cm}$\mbox{Distillation of Gaussian Einstein-Podolsky-Rosen steering with noiseless linear amplification}$}
\author{Yang Liu$^{\ddagger}$}
\address{State Key Laboratory of Quantum Optics and Quantum Optics Devices, Institute of Opto-Electronics, Shanxi University, Taiyuan 030006, China}
\address{Collaborative Innovation Center of Extreme Optics, Shanxi University, Taiyuan 030006, China}
\author{Kaimin Zheng$^{\ddagger}$}
\address{National Laboratory of Solid State Microstructures, Key Laboratory of Intelligent Optical Sensing and Manipulation, College of Engineering and Applied Sciences, and Collaborative Innovation Center of Advanced Microstructures, Nanjing University, Nanjing 210093, China}
\author{Haijun Kang}
\address{State Key Laboratory of Quantum Optics and Quantum Optics Devices, Institute of Opto-Electronics, Shanxi University, Taiyuan 030006, China}
\address{Collaborative Innovation Center of Extreme Optics, Shanxi University, Taiyuan 030006, China}
\author{Dongmei Han}
\address{State Key Laboratory of Quantum Optics and Quantum Optics Devices, Institute of Opto-Electronics, Shanxi University, Taiyuan 030006, China}
\address{Collaborative Innovation Center of Extreme Optics, Shanxi University, Taiyuan 030006, China}
\author{Meihong Wang}
\address{State Key Laboratory of Quantum Optics and Quantum Optics Devices, Institute of Opto-Electronics, Shanxi University, Taiyuan 030006, China}
\address{Collaborative Innovation Center of Extreme Optics, Shanxi University, Taiyuan 030006, China}
\author{Lijian Zhang}
\email{lijian.zhang@nju.edu.cn}
\address{National Laboratory of Solid State Microstructures, Key Laboratory of Intelligent Optical Sensing and Manipulation, College of Engineering and Applied Sciences, and Collaborative Innovation Center of Advanced Microstructures, Nanjing University, Nanjing 210093, China}
\author{Xiaolong~Su}
\email{suxl@sxu.edu.cn}
\address{State Key Laboratory of Quantum Optics and Quantum Optics Devices, Institute of Opto-Electronics, Shanxi University, Taiyuan 030006, China}
\address{Collaborative Innovation Center of Extreme Optics, Shanxi University, Taiyuan 030006, China}
\author{Kunchi Peng}
\address{State Key Laboratory of Quantum Optics and Quantum Optics Devices, Institute of Opto-Electronics, Shanxi University, Taiyuan 030006, China}
\address{Collaborative Innovation Center of Extreme Optics, Shanxi University, Taiyuan 030006, China}

\begin{abstract}

Einstein-Podolsky-Rosen (EPR) steering is one of the most intriguing features of quantum mechanics and an important resource for quantum communication. For practical applications, it remains a challenge to protect EPR steering from decoherence due to its intrinsic difference from entanglement. Here, we experimentally demonstrate the distillation of Gaussian EPR steering and entanglement in lossy and noisy environments using measurement-based noiseless linear amplification. Different from entanglement distillation, the extension of steerable region happens in the distillation of EPR steering besides the enhancement of steerabilities. We demonstrate that the two-way or one-way steerable region is extended after the distillation of EPR steering when the NLA is implemented based on Bob's or Alice's measurement results. We also show that the NLA helps to extract secret key from insecure region in one-sided device-independent quantum key distribution with EPR steering. Our work paves the way for quantum communication exploiting EPR steering in practical quantum channels.

\end{abstract}

\maketitle

\section*{Introduction}

Early in 1935, Schr\"{o}dinger put forward the term `steering' to
describe the `spooky action-at-a-distance' phenomenon pointed out by
Einstein, Podolsky, and Rosen (EPR) in their famous paradox \cite{Schrodinger35,EPR35}, where local measurements on one subsystem can apparently adjust (steer) the state of another distant subsystem~\cite{Howard07PRL,ReidRMP,Eric09,cavalcanti17review,prxresource,SQT16_LiCM,He15,Adesso15,RMP2020}. Most importantly, EPR steering is an intermediate type of quantum correlation between entanglement and Bell nonlocality \cite{Howard07PRL}. The test of EPR steering is often implemented in the one-sided device-independent (1sDI) scenario where one of the two parties uses uncharacterized measurement device, which is different from the test of entanglement where two parties use well-characterized measurement devices. EPR steering is intrinsically asymmetric with respect to the two subsystems, and the steerability from one subsystem to the other maybe different from that of the reverse direction. In certain situations, the steerability may only exist for one direction which is called one-way steering \cite{OneWayNatPhot,LPK,OneWayGuo,deng,zhong,prl121,prl2018,SA}. Due to this intriguing feature, EPR steering has been identified as a unique physical resource for 1sDI quantum key distribution (QKD)~\cite%
{1sDIQKD_howard,CV-QKDexp,HowardOptica}, secure quantum teleportation~\cite{SQT15}, and subchannel discrimination \cite{subchannel}.

\begin{figure*}[tbp]
\begin{center}
\includegraphics[width=140mm]{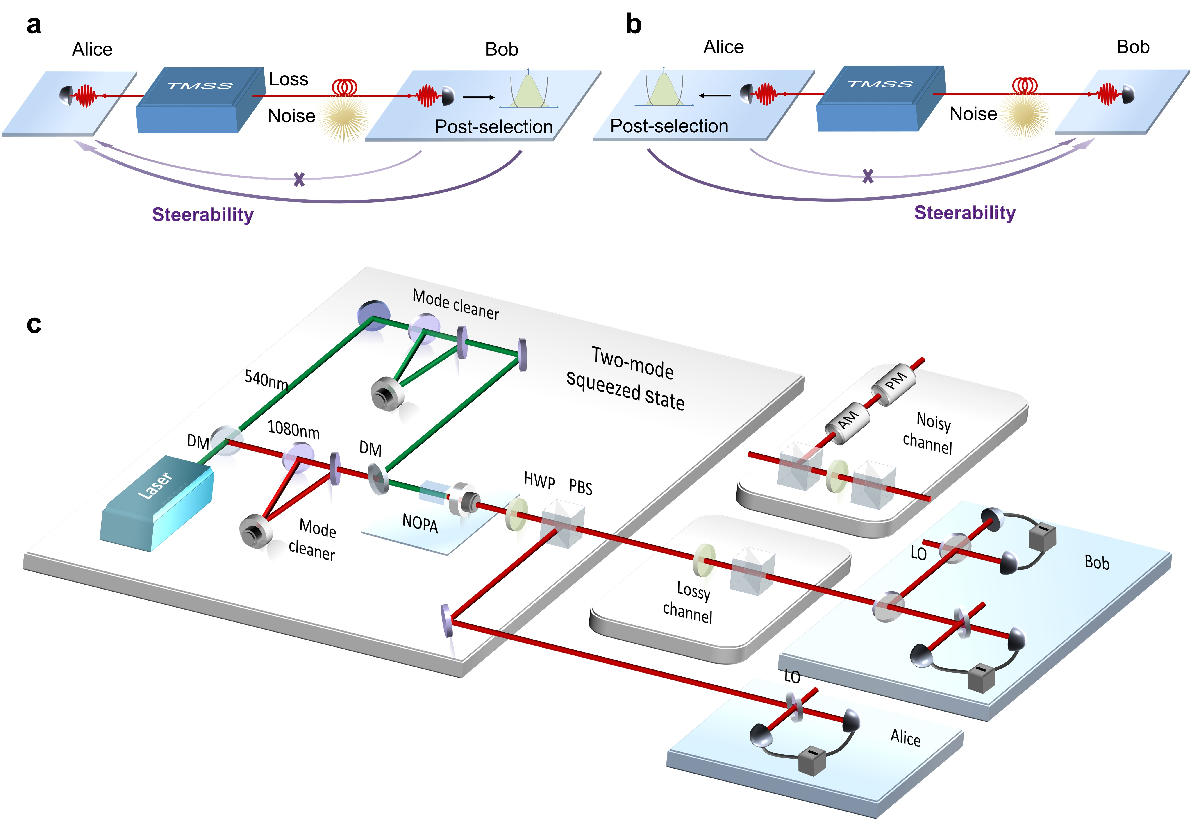}
\end{center}
\caption{\textbf{Schematic and experimental setup of the distillation. a} The measurement-based NLA based on Bob's measurement results. One mode of the two-mode squeezed state (TMSS) is transmitted to Bob through a lossy or noisy channel (remote mode), and the other mode is kept by Alice (local mode). After the NLA based on Bob's measurement results, the disappeared steerability from Bob to Alice is recovered. \textbf{b} The measurement-based NLA based on Alice's measurement results. In the case of noisy channel, the disappeared steerability from Alice to Bob is recovered after the NLA based on Alice's measurement results. \textbf{c} Experimental setup for the distillation when the NLA is implemented based on Bob's measurement results. The lossy channel is simulated by a half-wave plate (HWP) and a polarization beam-splitter (PBS). The noisy channel is modeled by combining the transmitted mode and another auxiliary beam modulated by electro-optic modulators (EOMs) on a PBS followed by a HWP and a PBS. The added excess noise is Gaussian noise with zero mean value. AM: amplitude modulator; PM: phase modulator; DM: dichroic mirror; LO: local oscillator.}
\end{figure*}

One of the obstacles in quantum information is decoherence, which unavoidably reduce the quantum property of quantum state and deteriorate the performance of quantum information processing. Therefore it is crucial to protect the quantum resource against decoherence. For quantum entanglement, this goal can be achieved with entanglement distillation~\cite{PRLD,nature,subnp2010,subprl,xiang,NLA,npaeu}, which recovers or increases entanglement from noisy entangled states. For Gaussian entanglement, there have been various proposals for entanglement distillation including photon subtraction \cite{subnp2010,subprl} and noiseless linear amplification (NLA) \cite{pra91,pra86,NLA,iop,prlff,npaz,xiang,npaeu}. Recently, it has been shown that measurement-based NLA, which is realized by post-selecting measured data with a designed filter function, is equivalent to physical NLA and it is used to protect Gaussian entanglement in a lossy channel~\cite{NLA}. However, the distillation of entanglement in a noisy environment still remains a challenge up to now.

In contrast to entanglement, the steerability of two directions between the subsystems decreases asymmetrically in a decoherent environment. In particular, the two-way steering may turn into one-way in the pure lossy channel and may disappear completely with excess noise \cite{zhong}. Therefore protecting EPR steering against loss and noise is an urgent but also more complex task. Very recently, the non-Markovian environment has been used to revive the disappeared Gaussian EPR steering in a noisy channel \cite{npjqi}, but the revived steerability with a correlated noisy channel cannot exceed the initial steerability. Only recently the distillation of EPR steering has drawn the attention and the preliminary results confirm the difference from entanglement distillation \cite{distilsteering}. However, the methods to distill EPR steering have been largely unexplored. In addition, how the directions of steering are affected by distillation remains an open question.

In this work, we experimentally demonstrate the distillation of Gaussian EPR steering using measurement-based NLA. At first, we implement and compare the distillation of Gaussian entanglement and EPR steering in a pure lossy channel, where only vacuum noise exists in the channel, via performing the NLA based on the measurement results of the remote mode, which is transmitted to Bob through a lossy quantum channel. The distilled EPR steering is enhanced for both directions and two-way EPR steering is recovered from one-way EPR steering in certain region of loss, which is different from that of entanglement. Then, we demonstrate the distillation of Gaussian entanglement and EPR steering in a noisy channel, where an additional thermal noise exists in the channel, with the NLA based on Alice's and Bob's measurement results, respectively. We show that the one-way EPR steering can be recovered from non-steerable region in a noisy environment by implementing the NLA based on Alice's measurement results, which can not be observed in the entanglement distillation. Our results confirm that measurement-based NLA may also find applications in noisy environment, which have not been investigated previously. In terms of application, we find that the secret key in continuous variable (CV) 1sDI QKD can also be distilled with the measurement-based NLA from an insecure regime.

\section*{Results}

\begin{figure*}[tbp]
\begin{center}
\includegraphics[width=160mm]{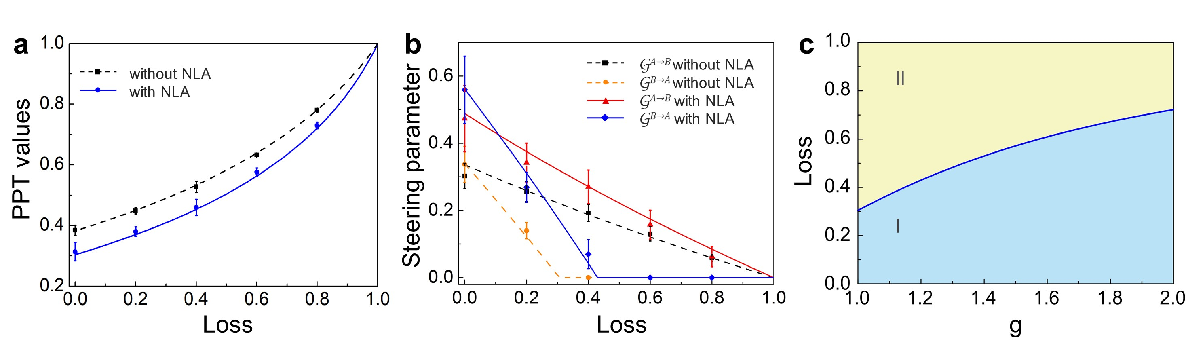}
\end{center}
\caption{\textbf{Results for the distillation in a lossy channel.} The PPT values \textbf{a} and the Gaussian EPR steering \textbf{b} with and without the NLA based on Bob's measurement results, respectively. \textbf{c} The EPR steerable regions parameterized by loss and gain with the NLA based on Bob's measurement results in a lossy channel. The blue and yellow regions are the two-way steerable region (I) and one-way steerable region (II), respectively. Error bars correspond to one standard deviation from statistical data.}
\end{figure*}

\noindent \textbf{The principle and experimental setup}

Since EPR steering is a phenomenon in the 1sDI scenario, we implement measurement-based NLA based on the result of well-characterized measurement device. As shown in Fig. 1a, when well-characterized measurement device is used by Bob we implement the NLA based on Bob's measurement results. Bob performs heterodyne measurement on his received state and decides whether to keep the measurement result $\beta $\ with the
acceptance probability $P_{acc}\left( \beta \right)$, which is given by
\begin{equation}
P_{acc}\left( \beta \right) =\left\{
\begin{array}{cr}
e^{\left( 1-g^{-2}\right) \left( \left\vert \beta \right\vert
^{2}-\left\vert \beta _{c}\right\vert ^{2}\right)} ,&{\left\vert \beta
\right\vert <\left\vert \beta _{c}\right\vert } \\
1,&{\left\vert \beta \right\vert \geq \left\vert \beta _{c}\right\vert }
\end{array}%
\right.   .
\end{equation}
After that, Bob announces his decision and Alice keeps or discards her measurement results accordingly. Here, the kept data is corresponding to the success of NLA. We note that the acceptance rate of the measurement-based NLA decreases along with the increase of cutoff $\left\vert \beta_{c}\right\vert$ while the fidelity of truncated filter with the ideal NLA increases with $\left\vert \beta_{c}\right\vert$. The optimal cutoff $\left\vert \beta_{c}\right\vert$ depends on the input state and the amplification gain $g$, which is determined numerically (see Appendix B). For the NLA based on Alice's measurement results, which corresponds to the case that well-characterized measurement device is used by Alice, the roles of Alice and Bob are swapped [Fig. 1b].

In the experiment for the distillation of Gaussian entanglement and EPR steering with the NLA based on Bob's measurement results,  Alice measures either the amplitude $\hat{x}=\hat{a}+\hat{a}^{\dag}$ or the phase $\hat{p}=-i(\hat{a}-\hat{a}^{\dag})$ quadrature of her state with homodyne detection, while Bob performs heterodyne detection on his state, which measures both quadratures simultaneously (Fig. 1c). A two-mode squeezed state (TMSS)  with $-$4.2 dB squeezing and 7.3 dB antisqueezing in time-domain is prepared by a nondegenerate optical parametric amplifier (NOPA) operating at deamplification status, which consists of an $\alpha $-cut type-II KTiOPO4 (KTP) crystal and an output coupling mirror (see details in ``Methods''). In order to measure the quantum noise of the TMSS in time-domain, the output signals of the homodyne detectors are mixed with a local reference signal of 3 MHz and then filtered by low-pass filters with bandwidth of 30 kHz and amplified 500 times (Low noise preamplifier, SRS, SR560). The output signals of the preamplifiers are recorded by a digital storage oscilloscope simultaneously. A sample size of $10^{8}$ data points is used for all quadrature measurements with sampling rate of
500 KS/s. 

\begin{figure*}[htbp]
\begin{center}
\includegraphics[width=160mm]{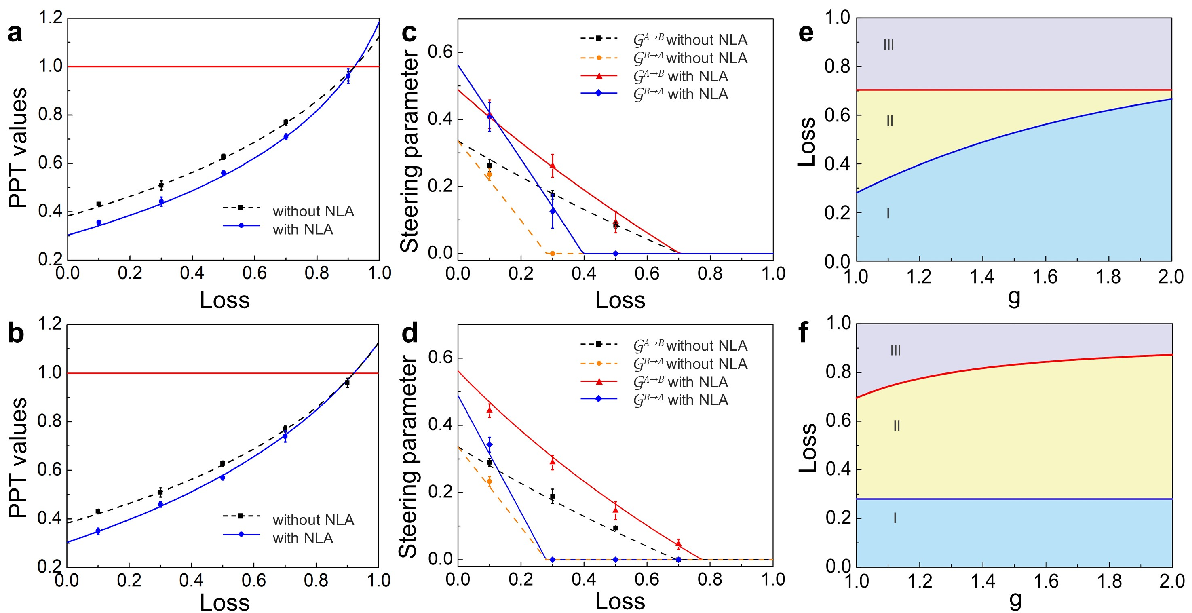}
\end{center}
\caption{\textbf{Results for the distillation in a noisy channel in which the excess noise is taken as 0.12 times of vacuum noise.} The PPT values with and without the NLA based on Bob's \textbf{a} and Alice's \textbf{b} measurement results, respectively. Gaussian EPR steering with and without the NLA based on Bob's \textbf{c} and Alice's \textbf{d} measurement results, respectively. The EPR steerable regions parameterized by loss and gain with the NLA based on Bob's \textbf{e} and Alice's \textbf{f} measurement results, respectively. The blue, yellow and gray regions are the two-way steerable region (I), one-way steerable region (II), and non-steerable region (III), respectively. Error bars correspond to one standard deviation from statistical data.}
\end{figure*}

\noindent \textbf{The criteria of Gaussian entanglement and EPR steering}

The properties of a ($n+m$)-mode Gaussian state $\rho _{AB}$ can be determined by its covariance matrix
\begin{equation}  \label{eq:CM}
\sigma _{AB}=\left(
\begin{array}{cc}
A & C \\
C^{\top } & B%
\end{array}
\right) ,
\end{equation}
with elements $\sigma _{ij}=\langle \hat{\xi}_{i}\hat{\xi}_{j}+\hat{\xi}_{j}
\hat{\xi}_{i}\rangle /2-\langle \hat{\xi}_{i}\rangle \langle \hat{\xi}%
_{j}\rangle $, where $\hat{\xi}\equiv (\hat{x}_{1}^{A},\hat{p}_{1}^{A},...,\hat{x}_{n}^{A},\hat{p}%
_{n}^{A},\hat{x}_{1}^{B},\hat{p}_{1}^{B},...,\hat{x}_{m}^{B},\hat{p}%
_{m}^{B})^{\intercal }$  is the vector of the amplitude and phase
quadratures of optical modes. The submatrices $A$ corresponds to Alice's state and $B$ corresponds to Bob's state, respectively.

The positive partial transposition (PPT) criterion \cite{PPT} is a necessary and sufficient criterion for entanglement of a two-mode Gaussian state. The PPT value used to quantify the entanglement is the smallest sympletic eigenvalue $\mu$ of the partially transposed matrix $\sigma _{AB}^{\top_k}=T_k\sigma _{AB}T_k^{\top}$, where $\top_k$ represents the partial transposition with respect to mode $k$ $(k=1,2,...,n+m)$ and $T_k$~is a unit diagonal matrix except for~$T_{2k,~2k}=-1$\cite{PPT}.
If the smallest symplectic eigenvalue $\mu$ is below 1, the state is inseparable (entangled).
For the case of the TMSS, the PPT value is expressed as
\begin{equation}
\mu=\frac{1}{\sqrt{2}}\sqrt{\Gamma-\sqrt{\Gamma^2-4~\text{det} \sigma_{AB}}}
\end{equation}
where~$\Gamma=\text{det} {A}+\text{det} {B}-2~\text{det} C$.

The steerability of Bob by Alice ($A\rightarrow B$) for a ($n+m$)-mode
Gaussian state can be quantified by~\cite{Adesso15}
\begin{equation}  \label{eqn:parameter}
\mathcal{G}^{A\rightarrow B}(\sigma _{AB})=\max \bigg\{0, \underset{j:\bar{%
\nu}_{j}^{AB\backslash A}<1}{-\sum }\ln (\bar{\nu}_{j}^{AB\backslash A})%
\bigg\},
\end{equation}
where $\bar{\nu}_{j}^{AB\backslash A}$ $(j=1,...,m_B)$ are the symplectic
eigenvalues of $\bar{\sigma}_{AB\backslash A}=B-C^{\mathsf{T}}A^{-1}C$,
derived from the Schur complement of $A$ in\ the covariance matrix $\sigma
_{AB}$. The steerability of Alice by Bob [$%
\mathcal{G}^{B\rightarrow A}(\sigma _{AB})$] can be obtained by swapping the
roles of $A$ and $B$.

\begin{figure*}[tbp]
\begin{center}
\includegraphics[width=165mm]{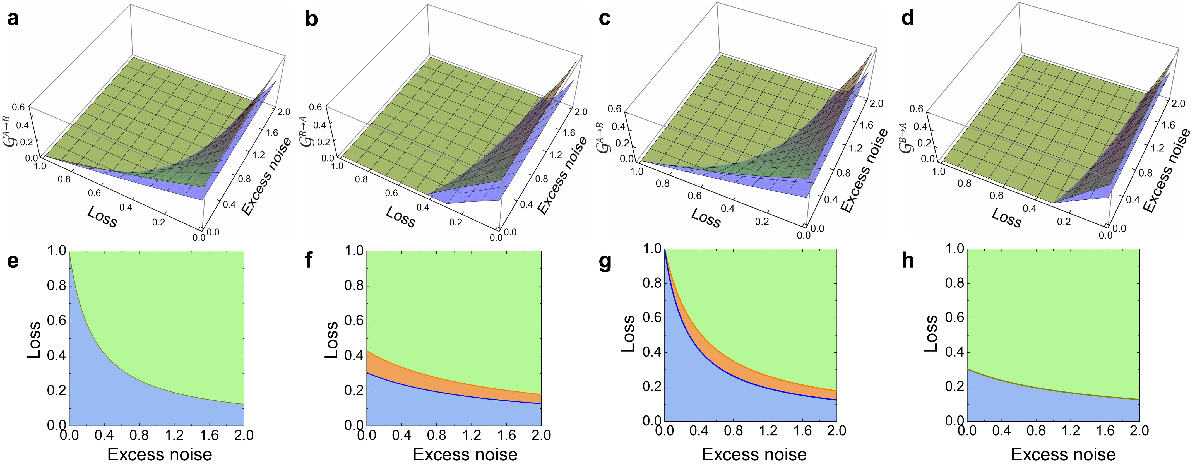}
\end{center}
\caption{\textbf{Results for the distillation of Gaussian EPR steering with different excess noise levels.} The steerabilities of $\mathcal{G}^{A\rightarrow B}$ \textbf{a} and $\mathcal{G}^{B\rightarrow A}$ \textbf{b} parameterized by loss and excess noise with (green) and without (blue) the NLA based on Bob's measurement results with gain $g=1.2$, respectively. The steerabilities of $\mathcal{G}^{A\rightarrow B}$ \textbf{c} and $\mathcal{G}^{B\rightarrow A}$ \textbf{d} parameterized by loss and excess noise with (green) and without (blue) the NLA based on Alice's measurement results with gain $g=1.2$, respectively. \textbf{e} to \textbf{h} are the EPR steerable regions parameterized by loss and excess noise, which are obtained from the projections of \textbf{a} to \textbf{d}, respectively. Green region: non-steerable region; Blue region: the steerable region before NLA; Orange region: the steerable region extended after the NLA.}
\end{figure*}

\noindent \textbf{Distillation results in a lossy channel}

From the experimentally reconstructed covariance matrix of the TMSS, the entanglement and EPR steering are obtained according to Eqs. (3) and (4), respectively. The result of Gaussian entanglement distillation in a lossy channel is shown in Fig. 2a, where the dependence of PPT values on the loss in quantum channel is presented. The entanglement between Alice and Bob decreases with the increase of loss but it is robust against loss since it disappears only when the loss reaches 1. After performing the NLA based on Bob's measurement results with gain $g=1.2$, the entanglement is enhanced, which is confirmed by the reduction of the PPT value.

The dependence of EPR steering on the loss is shown in Fig. 2b. It is obvious that the steerabilities for both directions decrease with the increase of the loss. When the loss is larger than 0.32, the steerability from Bob to Alice $\mathcal{G}^{B\rightarrow A}$ disappears, but the steerability from Alice to Bob $\mathcal{G}^{A\rightarrow B}$ is robust against loss in a lossy channel. This phenomenon confirms the unique property of one-way EPR steering, which is different from entanglement. After performing the NLA based on Bob's measurement results with gain $g=1.2$, the steerability for both directions are enhanced. Especially, the steerable range of $\mathcal{G}^{B\rightarrow A}$ is extended from 0.32 to 0.43, where the two-way steering is recovered in this region. The results confirm the feasibility of distilling Gaussian EPR steering in a lossy environment by using measurement-based NLA.

Figure 2c shows the EPR steerable regions parameterized by loss and gain with the NLA based on Bob's measurement results in a lossy environment. The two-way steerable region increases with the increase of gain in the NLA, while the one-way steerable region decreases. Comparing the results in Fig. 2a with Fig. 2b and 2c, we can see that the distillation result for EPR steering is different from that of entanglement, which comes from the fact of asymmetric property of EPR steering. When the NLA based on Alice's measurement results is implemented in a lossy channel, both entanglement and EPR steering can also be enhanced but the steerable region can not be extended  (see Appendix D).

\noindent \textbf{Distillation results in a noisy channel}

Noise in quantum channels is another key restriction factor in quantum communication besides loss. We experimentally demonstrate the distillation of Gaussian entanglement and EPR steering in a noisy channel in two cases, where the NLA based on measurement results of remote mode (Bob) and local mode (Alice) are implemented respectively. In the case of distillation based on Alice's measurement results, Alice and Bob perform heterodyne and homodyne detections respectively with excess noise existing in quantum channel from the TMSS to Bob. In our experiment, the excess noise is taken as 0.12 times of vacuum noise whose variance is 1.

In the presence of excess noise, entanglement disappears when the loss is higher than 0.92 without the measurement-based NLA (black dash curves in Fig. 3a and 3b). After applying the NLA based on Bob's measurement results with gain $g=1.2$, the entanglement is enhanced in the entangled region, but the region cannot be extended by the NLA (Fig. 3a). After applying the NLA based on Alice's measurement results, similar results is observed (Fig. 3b). The maximum entanglement with zero loss and the crucial point of the death of entanglement are overlapped in Fig. 3a and 3b, but the distilled entanglement with the NLA based on Bob's measurement results is slightly stronger than that based on Alice's measurement results.

For the distillation of Gaussian EPR steering, as shown in Fig. 3c and 3d, the steerability of $\mathcal{G}^{A\rightarrow B}$ disappears when the loss is higher than 0.73 in a noisy channel without the measurement-based NLA. After applying the NLA based on Bob's measurement results with gain $g=1.2$, the steerability of both $\mathcal{G}^{A\rightarrow B}$ and $\mathcal{G}^{B\rightarrow A}$ are also enhanced (Fig. 3c). The steerable range of $\mathcal{G}^{B\rightarrow A}$ is extended from 0.28 to 0.40, while that of $\mathcal{G}^{A\rightarrow B}$ cannot be extended by the NLA. For the distillation based on Alice's measurement results with $g=1.2$, the steerability of both directions are enhanced after the NLA and the steerable range of  $\mathcal{G}^{A\rightarrow B}$ is extended, i.e., the one-way steering is recovered from non-steerable region in a certain extent, but that of $\mathcal{G}^{B\rightarrow A}$ cannot be extended (Fig. 3d). This result together with the result in Fig. 3c confirm that the NLA can protect the Gaussian EPR steering against noise.

The steerability of $\mathcal{G}^{A\rightarrow B}$ and $\mathcal{G}^{B\rightarrow A}$ are the same at loss $= 0$ without NLA as shown in Figs. 2b, 3c and 3d. This is because the initial state is symmetric, i.e., the submatrices of Alice's and Bob's states [$A$ and $B$ in Eq. (2)] are the same. In this case, $\mathcal{G}^{A\rightarrow B}$ and $\mathcal{G}^{B\rightarrow A}$ are the same according to Eq. (4). While $\mathcal{G}^{A\rightarrow B}$ and $\mathcal{G}^{B\rightarrow A}$ become different at loss $= 0$ after the NLA, as shown in Fig. 2b, 3c and 3d. This is because the state becomes asymmetric after the NLA, i.e., the submatrices of Alice's and Bob's states are different, since the initial state in our experiment is not a pure TMSS. Interestingly, we find that the steerability of $\mathcal{G}^{A\rightarrow B}$ and $\mathcal{G}^{B\rightarrow A}$ remain the same after the NLA at loss $= 0$ for the distillation of a pure TMSS because the state remains symmetric after the NLA (see Appendix C).

Figure 3e and 3f show the EPR steerable regions parameterized by loss and gain with the NLA based on Bob's and Alice's measurement results in a noisy channel, respectively. It is obvious that the two-way steerable region increases with the increase of gain in the NLA, and the non-steerable region is not affected by the gain when the NLA based on Bob's measurement results is applied. While in the NLA based on Alice's measurement results, the two-way EPR steerable region is not affected by the gain, but the non-steerable region is reduced with the gain.

\begin{figure*}[tbp]
\begin{center}
\includegraphics[width=120mm]{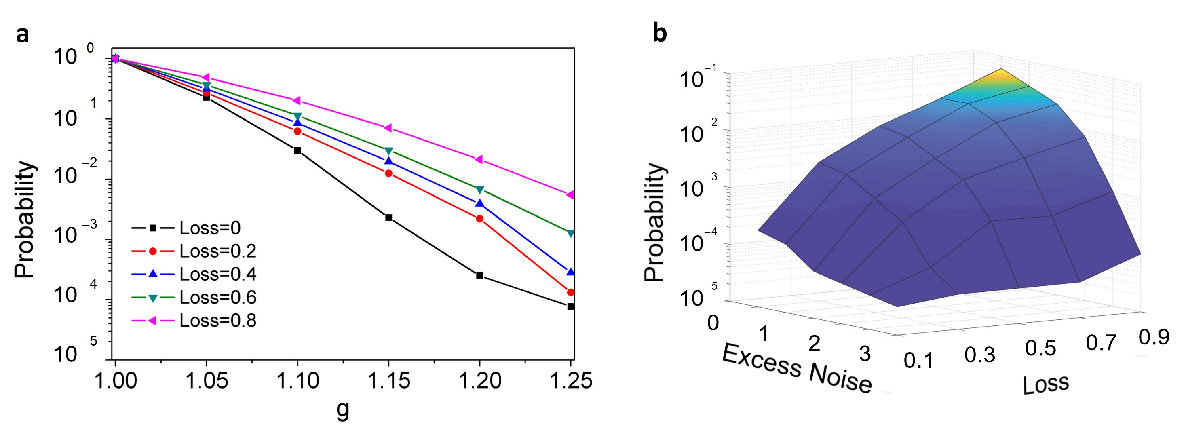}
\end{center}
\caption{\textbf{The probabilities of the success.} \textbf{a} The probability of success as a function of the $g$ for different losses in a lossy channel. \textbf{b} The dependence of probability of success on loss and excess noise with a certain amplification gain $g=1.2$ in a noisy channel.}
\end{figure*}

We also show the dependence of the distillation of Gaussian EPR steering on excess noise levels. The dependence of the steerabilities of $\mathcal{G}^{A\rightarrow B}$ and $\mathcal{G}^{B\rightarrow A}$ on loss and excess noise with and without the NLA implemented based on Bob's measurement results are shown in Fig. 4a and 4b, respectively. After the NLA with gain $g=1.2$, the steerabilities for both directions are enhanced with the excess noise up to 2 times of the vacuum noise. We show that both the steerable region of $\mathcal{G}^{A\rightarrow B}$ [Fig. 4e] and $\mathcal{G}^{B\rightarrow A}$ [Fig. 4f] decrease with the increase of excess noise. After the NLA, the steerable range of $\mathcal{G}^{B\rightarrow A}$ is extended, but that of $\mathcal{G}^{A\rightarrow B}$ cannot be extended, which is same to the result obtained in Fig. 3c where the excess noise is fixed to $0.12$. If Alice wants to enlarge the steerable range of $\mathcal{G}^{A\rightarrow B}$, it can be realized by performing the NLA based on Alice's measurement results. In this case, the steerable range of $\mathcal{G}^{A\rightarrow B}$ is extended [Fig. 4g], but that of $\mathcal{G}^{B\rightarrow A}$ cannot be extended [Fig. 4h] with the excess noise up to 2 times of the vacuum noise, which is different from the result of the NLA based on Bob's measurement results.

The probability of success of the NLA for different losses and gain in a lossy channel is shown in Fig. 5a. Firstly, we can see that the probability of success decreases with the increase of $g$ for a certain loss, which means that the NLA with larger gain coefficient is more difficult to achieve for the same initial state. Moreover, for a certain gain, the probability of success increases with the increase of loss. Please note that different optimal cutoff values shown in  Table 1 are chosen in this case. In case of noisy channel, the probability of success of the NLA for different losses and different excess noises with a certain gain $g=1.2$ is shown in Fig. 5b. For a certain excess noise, the probability of success increases with the increase of loss. For a certain loss, the probability decreases with the increase of excess noise. So the excess noise cannot be infinite.

\noindent \textbf{Application in 1sDI QKD}

As an example of application, we apply our scheme to the CV 1sDI QKD. The 1sDI QKD is a protocol that only one of the two measurement apparatus is trusted~\cite
{1sDIQKD_howard}. When Alice and Bob perform homodyne and heterodyne detection on their states respectively, it corresponds to the CV 1sDI QKD with Homodyne-Heterodyne measurements~\cite{HowardOptica}. In the case of reverse reconciliation, in which Bob sends corrections to Alice, the secret key rate for this CV 1sDI QKD protocol is bounded by~\cite{HowardOptica},

\vspace{-1.2cm}
\begin{align}
K^{\blacktriangleleft} &\geq S\left(X_{B}|E\right)-H\left(X_{B}|X_{A}\right) \nonumber \\
&\geq \text{log}_{2}{\frac{2}{e\sqrt{V_{P_{B}|P_{A}}V_{X_{B}|X_{A}}}}}
\end{align}
where $S\left(X_{B}|E\right)$ is the conditional von Neumann entropy of $X_{B}$ given $E$, $H\left(X_{B}|X_{A}\right)$ is the Shannon entropy of measurement strings of $X_{B}$ given $X_{A}$. It should be noted that $V_{P_{B}|P_{A}}$ and $V_{X_{B}|X_{A}}$ are the conditional variance of Bob's heterodyne measurement given Alice's homodyne measurement. The conditional variances can be calculated directly from the measurement results.
So the secret key rate for this 1sDI QKD can be obtained according to Eq. (5).

As shown in Fig. 6, without measurement-based NLA the minimum squeezing level to obtain the secret key is $-6$ dB, which is because the security of the CV 1sDI QKD with homodyne-heterodyne measurements raises the requirement of Gaussian EPR steering (see Appendix E). In the case of our experiment, there is no secret key for the TMSS with $-4.2$ dB squeezing and $7.3$ dB antisqueezing without the NLA, as shown by the blue curve at the point of $g=1$ in Fig. 6. When the measurement-based NLA is applied, the secret key can be extracted with $g>1.4$. Thus, the measurement-based NLA can be used to distill secret key in CV 1sDI QKD with Homodyne-Heterodyne measurements.

\section*{Discussion}

In this work, the EPR steering criterion for Gaussian states is applied to quantify the steerabilities before and after the distillation. This criterion is valid because the Gaussian states after the NLA remains Gaussian. In the experiment we ensure that the data after measurement-based NLA follows a Gaussian distribution with proper choice of the parameters. There are other distillation protocols for entanglement, for example, with photon subtraction \cite{subnp2010,subprl}  and quantum catalysis \cite{npaeu}. However, the states after these operations are usually non-Gaussian. Up to now, there is still no criterion to rigorously quantify the steerabilities of non-Gaussian states. Therefore whether these distillation protocols also work for EPR steering remains to be explored.

In summary, we experimentally compare the distillation of Gaussian EPR steering and entanglement with the measurement-based NLA in both lossy and noisy environments. We observe that the one-way and two-way steerable regions are changed by the distillation of Gaussian EPR steering, which cannot occur in the entanglement distillation. Comparing with the entanglement distillation demonstrated in Ref. \cite{NLA}, where the measurement-based NLA based on Bob's measurement results is used to distill Gaussian entanglement in a lossy channel, we demonstrate the performance of measurement-based NLA in the presence of channel noise in addition to the lossy-only channel in our experiment. Most importantly, we show that the distillation results of Gaussian EPR steering are different when the NLA based on Alice's or Bob's measurement results.

Our results confirm the feasibility of protecting Gaussian EPR steering in a decoherent environment using measurement-based NLA. We also show that the distillation of Gaussian EPR steering with measurement-based NLA is helpful to distill secret key in the 1sDI QKD. Our work thus makes an essential step for applying EPR steering in improving fidelity of secure quantum teleportation and key rates in 1sDI QKD over practical quantum channels.

\section*{METHODS}

\noindent \textbf{Details of the experiment}

\begin{figure}[tbp]
\begin{center}
\includegraphics[width=80mm]{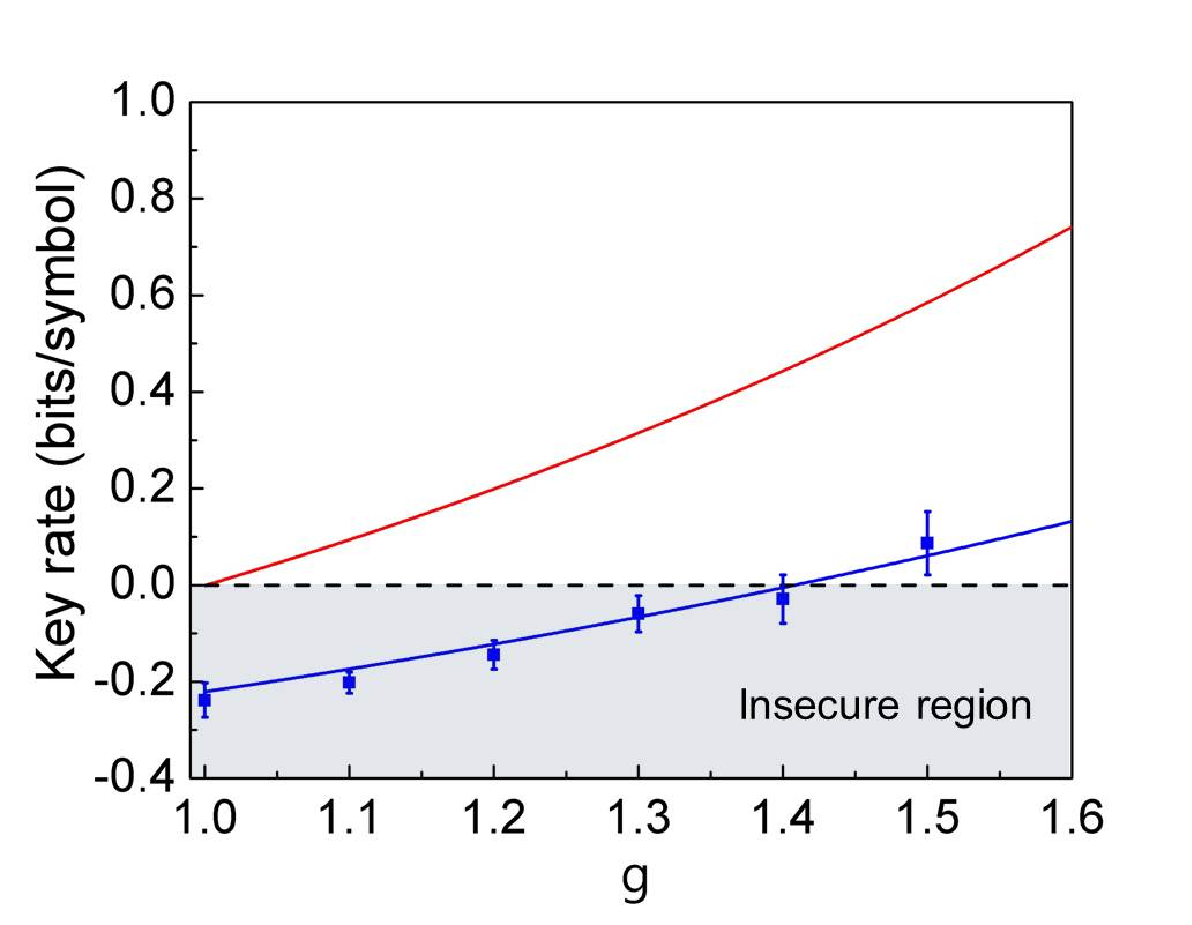}
\end{center}
\caption{\textbf{The secret key rate for 1sDI QKD with continuous variables in the case of reverse reconciliation.} The red curve represents the theoretical key rate given by a pure TMSS with $-6$ dB squeezing. The blue curve represents the key rate given by the initial state with $-4.2$ dB squeezing and $7.3$ dB antisqueezing. The cutoff is selected as $\beta _{c}=4.5$. Error bars correspond to one standard deviation from statistical data. The region with negative key rate is the insecure region.}
\end{figure}

Two mode cleaners are inserted between the laser source and the NOPAs to filter noise and higher order spatial modes of the laser beams at $540$ nm and $1080$ nm. The fundamental wave at $1080$ nm wavelength is used for the injected signals of the NOPAs and the local oscillators for the homodyne detectors. The second-harmonic wave at $540$ nm wavelength serves as pump field of the NOPAs, in which a pair of signal and idler modes with orthogonal polarizations at $1080$ nm are generated through an intracavity frequency-down-conversion process.

The NOPA is in a semimonolithic structure, where the front face of KTP crystal is coated to be used for the input coupler and the concave mirror serves as the output coupler of squeezed states. The transmittances of the front face of KTP crystal at $540$ nm and $1080$ nm are $40\%$ and $0.04\%$, respectively. The end-face of KTP is antireflection coated for both $1080$ nm and $540$ nm. The transmittances of output coupler at $540$ nm and $1080$ nm are $0.5\%$ and $12.5\%$, respectively. When a NOPA is operating at deamplification status (the relative phase between injected signal and pump beam is locked to~${(2n+1)\pi}$), the coupled modes are a TMSS with the anticorrelated amplitude quadrature and correlated phase quadrature.

To reconstruct the covariance matrix of the output state, the variances and the cross correlations of the amplitude or phase quadratures are obtained by simultaneously measuring the amplitude or phase quadratures of two modes of the TMSS in time domain. The diagonal elements of the covariance matrix  are the variances of the amplitude and phase quadratures $\bigtriangleup ^{2}(\hat{\xi}_{i})$, and the nondiagonal elements are the covariances of the amplitude or phase quadratures which are calculated via the measured variances \cite{cov}
\begin{equation}
\sigma \left(\hat{\xi}_{i},\hat{\xi}_{j}\right)=[\bigtriangleup ^{2}(\hat{\xi}_{i}+\hat{\xi}_{j})-\bigtriangleup ^{2}(\hat{\xi}_{i})-\bigtriangleup ^{2}(\hat{\xi}_{j})]/2
\end{equation}
\begin{equation}
\sigma \left(\hat{\xi}_{i},\hat{\xi}_{j}\right)=-[\bigtriangleup ^{2}(\hat{\xi}_{i}-\hat{\xi}_{j})-\bigtriangleup ^{2}(\hat{\xi}_{i})-\bigtriangleup ^{2}(\hat{\xi}_{j})]/2
\end{equation}
where $\bigtriangleup ^{2}(\hat{\xi}_{i}+\hat{\xi}_{j})$ and $\bigtriangleup ^{2}(\hat{\xi}_{i}-\hat{\xi}_{j})$ are the correlation variances of amplitude and phase quadratures, which can be obtained from the measured variances in time domain. Based on the reconstructed covariance matrix, the PPT values and steerablities can be quantified according to Eqs. (3) and (4), respectively.

This research was supported by National Natural Science Foundation of China
(Grants No. 11834010, No. 61975077 and No. 11690032), National Key R \& D Program of China (Grant  No. 2017YFA0303703). X. Su thanks the Fund for Shanxi ``1331 Project" Key Subjects Construction.

$^{\ddagger}$Y. Liu and K. Zheng contributed equally to this work.

\appendix	
	
\section{\label{sec:appendix1} Details of the theoretical calculation of measurement-based NLA}

\begin{figure*}[tbp]
\begin{center}
\includegraphics[width=120mm]{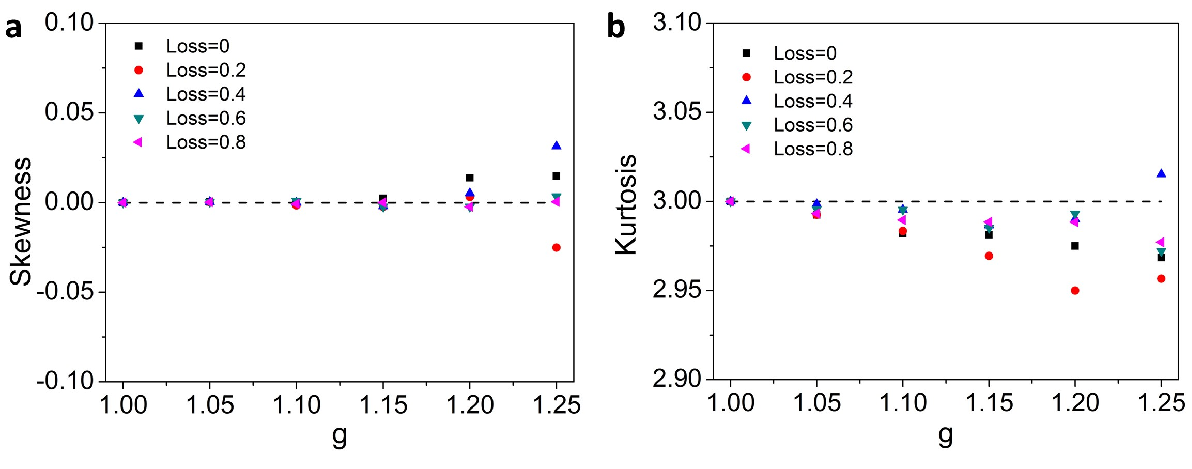}
\end{center}
\begin{center}
\caption{\textbf{The skewness and kurtosis.} \textbf{a} The skewness of the post-selected ensembles as a function of the $g$ for different losses. \textbf{b} The kurtosis of the post-selected ensembles as a function of the $g$ for different losses.}
\end{center}
\label{figs1}
\end{figure*}

The ideal NLA operation with gain $g$ can be described as $g^{
\hat{n}}$, where $\hat{n}=\hat{a}^{\dagger }\hat{a}$ is the
photon number operator. If Bob's state is $\rho _{B}$, after the NLA, the
probability of the heterodyne measurement outcome $\alpha $ is
\begin{align}
P_{g}\left( \alpha \right) &=\frac{1}{\pi }\left\langle \alpha \right\vert
g^{\hat{n}}\rho _{B}g^{\hat{n}}\left\vert \alpha \right\rangle \nonumber \\
&=\frac{1}{\pi }e^{\left( g^{2}-1\right) \left\vert \alpha \right\vert
^{2}}\left\langle g\alpha \right\vert \rho _{B}\left\vert g\alpha
\right\rangle ,
\label{eq:NLA}
\end{align}
where we used $g^{\hat{n}}\left\vert \alpha \right\rangle
=e^{\frac{\left( g^{2}-1\right) \left\vert \alpha \right\vert ^{2}}{2}%
}\left\vert g\alpha \right\rangle $. When $\rho _{B}$ is
measured with a heterodyne detection system directly, Bob will get the measurement outcome $\beta $ with
probability $P\left( \beta \right) =\frac{1}{\pi }\left\langle \beta
\right\vert \rho _{B}\left\vert \beta \right\rangle .$ If $\beta $ is
rescaled as $\alpha={\beta }/{g}$ and assigned to a probabilistic filter
function $P_{\beta}=e^{\left( 1-g^{-2}\right) \left\vert \beta \right\vert ^{2}}$,
Bob's measurement outcome follows the same distribution given by Eq. (\ref{eq:NLA}) and thus emulates the operation $g^{
\hat{n}}$.
Notably, the probabilistic filter function is always greater than 1 for $g>1$. So it is not a legitimate weighting probability which means that it is
impossible to implement a ideal NLA. To implement a good approximation to
the ideal NLA $g^{\hat{n}}$, we can introduce a finite cutoff $\left\vert \beta
_{c}\right\vert$ and renormalize the probabilistic filter function $P_{\beta}$ with respect to $P_{\beta_c}$, which gives the acceptance
probability $P_{acc}\left( \beta \right)$ in Eq. (1) in the main text.

For a two-mode state, its density operator can be expressed as the
following Weyl representation, i.e.,
\begin{equation}
\hat{\rho} _{in}=\int \frac{d^{2}\alpha d^{2}\beta }{\pi ^{2}}\chi _{in}\left(
\alpha ,\beta \right) D\left( -\alpha \right) D\left( -\beta \right) ,
\label{eqs1}
\end{equation}%
where $\chi _{in}$ and $D\left( \alpha \right) =\exp \left \{ \alpha \hat{a}^{\dag
}-\alpha ^{\ast }\hat{a}\right \} $ are the characteristic function
corresponding to $\rho _{in}$ and the displacement operator. When each mode of $\rho _{in}$ goes through a noiseless amplifier, denoted as $g_{1}^{\hat{a}^{\dag }\hat{a}}$
and $g_{2}^{\hat{b}^{\dag }\hat{b}}$,  the output state can be shown as%
\begin{align}
\hat{\rho} _{out}& =Ng_{2}^{\hat{b}^{\dag }\hat{b}}g_{1}^{\hat{a}^{\dag }\hat{a}}\rho _{in}g_{1}^{\hat{a}^{\dag
}\hat{a}}g_{2}^{\hat{b}^{\dag }\hat{b}}  \nonumber \\
& =N\int \frac{d^{2}\alpha d^{2}\beta }{\pi ^{2}}\chi _{in}\left( \alpha
,\beta \right) g_{2}^{\hat{b}^{\dag }\hat{b}}g_{1}^{\hat{a}^{\dag }\hat{a}}D_{a}\left( -\alpha
\right) D_{b}\left( -\beta \right) g_{1}^{\hat{a}^{\dag }\hat{a}}g_{2}^{\hat{b}^{\dag }\hat{b}},\label{eqs2}
\end{align}%
where $N$ is the normalization factor, which is determined by tr$\hat{\rho}
_{out}=1$. The characteristic function of the amplified state is given by%
\begin{align}
& \chi _{out}\left( \bar{\alpha},\bar{\beta}\right)   \nonumber \\
& =N\int \frac{d^{2}\alpha d^{2}\beta }{\pi ^{2}}\chi _{in}\left( \alpha
,\beta \right)  \nonumber \\
& \mathtt{tr}\left[ g_{2}^{\hat{b}^{\dag }\hat{b}}g_{1}^{\hat{a}^{\dag }\hat{a}}D_{a}\left( -\alpha
\right) D_{b}\left( -\beta \right) g_{1}^{\hat{a}^{\dag }\hat{a}}g_{2}^{\hat{b}^{\dag
}\hat{b}}D_{a}\left( \bar{\alpha}\right) D_{b}\left( \bar{\beta}\right) \right]
\nonumber \\
& =\frac{N}{\left( 1-g_{1}^{2}\right) \left( 1-g_{2}^{2}\right) }\int \frac{%
d^{2}\alpha d^{2}\beta }{\pi ^{2}}\chi _{in}\left( \alpha ,\beta \right)
\nonumber \\
& \times \exp \left \{ A\left( \left \vert \alpha \right \vert ^{2}+\left \vert
\bar{\alpha}\right \vert ^{2}\right) +C\left( \bar{\alpha}\alpha ^{\ast }+%
\bar{\alpha}^{\ast }\alpha \right) \right \}   \nonumber \\
& \times \exp \left \{ B\left( \left \vert \beta \right \vert ^{2}+\left \vert
\bar{\beta}\right \vert ^{2}\right) +D\left( \bar{\beta}\beta ^{\ast }+\bar{%
\beta}^{\ast }\beta \right) \right \} ,  \label{eqs3}
\end{align}%
where $A=\frac{g_{1}^{2}+1}{2\left( g_{1}^{2}-1\right)},~C=\frac{g_{1}}{%
1-g_{1}^{2}},~B=\frac{g_{2}^{2}+1}{2\left( g_{2}^{2}-1\right)},~D=\frac{g_{2}%
}{1-g_{2}^{2}}$. For simplification, taking $\alpha =x_{1}+ip_{1}%
,~\beta =x_{2}+ip_{2},~\bar{\alpha}=\bar{x}_{1}+i\bar{p}_{1},~\bar{\beta}=\bar{x}_{2}+i\bar{p}_{2}$, and $X=\left(x_{1},p_{1},x_{2},p_{2}\right) ^{T},~\bar{X}=\left( \bar{x}_{1},\bar{p}_{1},%
\bar{x}_{2}, \bar{p}_{2}\right) ^{T}$, then Eq. (\ref{eqs3}) becomes%
\begin{align}
\chi _{out}\left( \bar{X}\right) & =\frac{N}{\left( 1-g_{1}^{2}\right)
\left( 1-g_{2}^{2}\right) }\exp \left \{ \bar{X}^{T}G_{1}\bar{X}\right \}
\nonumber \\
& \times \int \frac{d^{4}X}{\pi ^{2}}\chi _{in}\left( X\right) \exp \left \{
X^{T}G_{1}X+\bar{X}^{T}G_{2}X\right \} ,  \label{eqs4}
\end{align}%
where
\begin{equation}
G_{1}=\left(
\begin{array}{cccc}
A & 0 & 0 & 0 \\
0 & A & 0 & 0 \\
0 & 0 & B & 0 \\
0 & 0 & 0 & B%
\end{array}%
\right) ,G_{2}=\left(
\begin{array}{cccc}
2C & 0 & 0 & 0 \\
0 & 2C & 0 & 0 \\
0 & 0 & 2D & 0 \\
0 & 0 & 0 & 2D%
\end{array}%
\right) .  \label{eqs5}
\end{equation}
For any Gaussian state, with mean $d$
and covariance matrix $\sigma _{AB}$, their characteristic function can be expressed as \cite{RMP}
\begin{equation}
\chi _{in}\left( X\right) =\exp \left \{ -\frac{1}{2}X^{T}\left( \Omega
\sigma _{AB}\Omega ^{T}\right) X-i\left( \Omega d\right) ^{T}X\right \} ,  \label{eqs6} \\
\end{equation}%
where $\Omega =\left(
\begin{array}{cccc}
0 & 1 & 0 & 0 \\
-1 & 0 & 0 & 0 \\
0 & 0 & 0 & 1 \\
0 & 0 & -1 & 0%
\end{array}%
\right) .$ Then substituting Eq. (\ref{eqs6}) into Eq. (\ref{eqs4}) and using the integration
formula%
\begin{equation}
\int d^{n}X\exp \left \{ -\frac{1}{2}X^{T}MX+X^{T}v\right \}
=\frac{\pi ^{n/2}}{\sqrt{|M|}}\exp \left \{ \frac{1}{2}v^{T}M^{-1}v\right \},
\label{s28}
\end{equation}%
we have
\begin{align}
& \chi _{out}\left( \bar{X}\right)  \nonumber \nonumber \\
& \rightarrow
\exp \left \{ -\frac{1}{2}\bar{X}^{T}\left[ G_{2}\left( 2G_{1}-\Omega
\sigma _{AB}\Omega ^{T}\right) ^{-1}G_{2}-2G_{1}\right] \bar{X}\right \}  \nonumber \\
& \exp \left \{
\frac{1}{2}\left( G_{2}\bar{X}\right) ^{T}\left( \Omega \sigma _{AB}\Omega
^{T}-2G_{1}\right) ^{-1}\left( -i\left( \Omega d\right) \right)
\right \}  \nonumber \\
& \exp \left \{ \frac{1}{2}\left( -i\left( \Omega d\right) \right)
^{T}\left( \Omega \sigma _{AB}\Omega ^{T}-2G_{1}\right) ^{-1}\left( G_{2}\bar{X}-i\left( \Omega
d\right) \right) \right \},
\end{align}%
which indicates that the covariance matrix of the amplified state is
\begin{align}
\sigma _{nla}& =\Omega ^{-1}\left[ G_{2}\left( 2G_{1}-\Omega \sigma _{AB}\Omega ^{T}\right)
^{-1}G_{2}-2G_{1}\right] \left( \Omega ^{T}\right) ^{-1}  \nonumber \\
& =G_{2}\left( 2G_{1}-\sigma _{AB}\right) ^{-1}G_{2}-2G_{1}  \label{s29}
\end{align}%
where $\Omega ^{T}G_{1}\Omega =G_{1},~\Omega ^{T}G_{2}\Omega =G_{2},~\Omega
^{-1}=\Omega ^{T}.$  When only mode $\hat{B}$ of the two-mode state go through noiseless amplifier, we can take limit $g_{1}\rightarrow 1$.

\begin{table}[h!]
\centering
\caption{\textbf{Table 1 The optimal cutoff $\left\vert \beta_{c}\right\vert$ at different loss and $g$}}
\begin{tabular}{cccccc}
\hline
 & g=1.05 & g=1.10 & g=1.15 & g=1.20 & g=1.25 \\
\hline
Loss=0.0 &  4.25 & 4.75 & 5.25 & 5.50 & 6.00 \\

Loss=0.2 &  4.00 & 4.25 & 4.50 & 4.75 & 5.25 \\

Loss=0.4 &  3.75 & 4.00 & 4.25 & 4.50 & 5.00 \\

Loss=0.6 &  3.50 & 3.75 & 4.00 & 4.25 & 4.50 \\

Loss=0.8 &  3.00 & 3.25 & 3.50 & 3.75 & 4.00\\
\hline
\end{tabular}
  \label{tab:shape-functions}
\end{table}

\section{\label{sec:appendix2}The selection of cutoff value}

\begin{figure*}[tbp]
\begin{center}
\includegraphics[width=120mm]{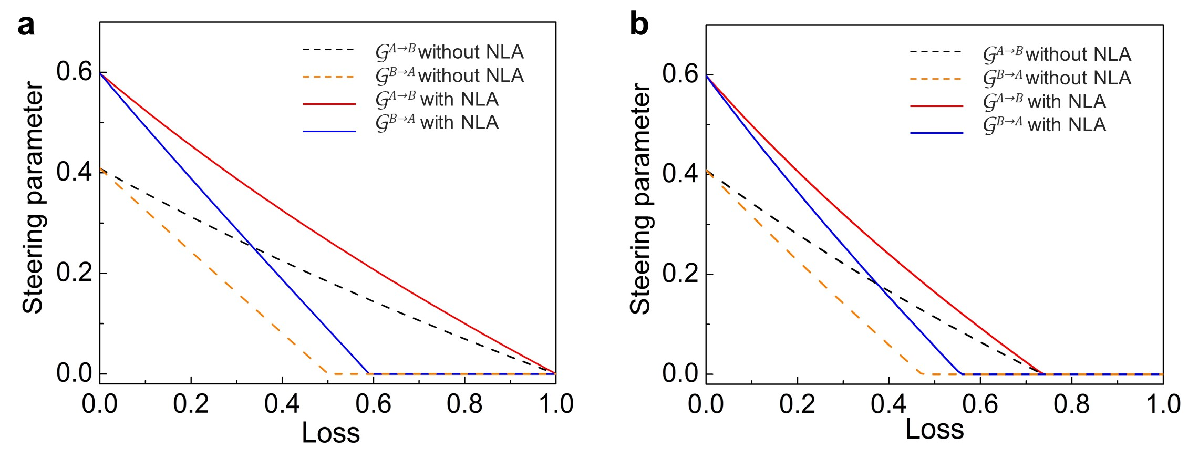}
\end{center}
\caption{\textbf{The distillation of Gaussian EPR steering for a pure TMSS.} The dependence of Gaussian EPR steering on loss for a pure two-mode squeezed state in a lossy channel \textbf{a} and in a noisy channel \textbf{b} with  and with out the NLA implemented based on Bob's measurement results. The black and yellow dashed lines are the steerability of $\mathcal{G}^{A\rightarrow B}$ and $\mathcal{G}^{B\rightarrow A}$ without NLA, respectively. The red and blue solid lines are the steerability of $\mathcal{G}^{A\rightarrow B}$ and $\mathcal{G}^{B\rightarrow A}$ with NLA, respectively.}
\label{figs3}
\end{figure*}

The optimal cutoff $\left\vert \beta_{c}\right\vert$ is selected to ensure high acceptance rate and high fidelity of the emulated NLA at the same time. However, we find that the post-selection rates is inversely proportional to cutoff $\left\vert \beta_{c}\right\vert$, and the fidelity of truncated filter with the ideal NLA is proportional to $\left\vert \beta_{c}\right\vert$. Here, we select the optimal $\left\vert \beta_{c}\right\vert$ according to the numerical simulation, in which the initial state is the same with that in the experiment. We choose the smallest $\left\vert \beta_{c}\right\vert$ to ensure the high fidelity of measurement-based NLA and ideal implementation $g^{\hat{n}}$ with two conditions: 1. the accepted data satisfy Gaussian distribution (skewness approaches to 0 and kurtosis approaches to 3); 2. the EPR steering calculated from the accepted data is close to that of the state after the ideal implementation $g^{\hat{n}}$. The optimal cutoff $\left\vert \beta_{c}\right\vert$ for different losses and $g$ are given in Table 1. With the increase of losses, the optimal cutoff value becomes smaller.

\section{\label{sec:appendix3}Distillation of EPR steering for a pure two-mode squeezed state}

The purity of a two-mode squeezed state can be described by \cite{purity}
\begin{equation}
\mathcal{P} =\frac{1}{\sqrt{\det (\sigma _{AB})}},
\end{equation}
where $\sigma _{AB}$ is the covariance matrix of the two-mode squeezed state. For a two-mode squeezed state with $-$4.2 dB squeezing and 4.2 dB antisqueezing, we have purity $\mathcal{P}_{p}=1$. The theoretical results for the distillation of EPR steering with a pure two-mode squeezed state are shown in Fig. 8. After measurement-based NLA based on Bob's measurement results with gain $g=1.2$ is implemented, the steerability for both directions are enhanced and the tolerance of $\mathcal{G}^{B\rightarrow A}$ on loss is extended in both lossy and noisy channels. Please note that the steerability from Bob to Alice $\mathcal{G}^{B\rightarrow A}$ is equal to $\mathcal{G}^{A\rightarrow B}$ when there is no loss and $\mathcal{G}^{B\rightarrow A}$ never surpasses $\mathcal{G}^{A\rightarrow B}$ for a pure two-mode squeezed state after NLA.

In our experiment, a two-mode squeezed state with $-$4.2 dB squeezing and 7.3 dB antisqueezing is prepared with purity $\mathcal{P}=0.55$, which is not pure. The impurity of two-mode squeezed state in the experiment is caused by losses in the experimental system. The resson for steerabilities $\mathcal{G}^{B\rightarrow A}$ and $\mathcal{G}^{A\rightarrow B}$ are different after the NLA in Figures 2b, 3c and 3d in the main text is that the state becomes asymmetric i.e., the submatrices of Alice's and Bob's states are different after the NLA for the impure TMSS. While this is not happened after the NLA for a pure TMSS since the state remains symmetric, as shown in Fig. 8.

\section{\label{sec:appendix4}The distillation in a lossy channel based on Alice's measurement results}

\begin{figure*}[tbp]
\begin{center}
\includegraphics[width=165mm]{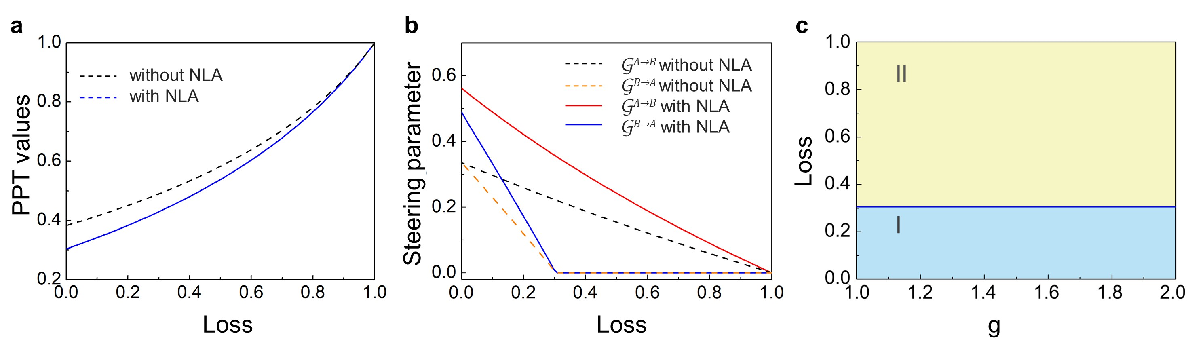}
\end{center}
\caption{\textbf{The distillation results in a lossy channel based on Alice's measurement results.} The PPT values \textbf{a} and the Gaussian EPR steering \textbf{b} with and without the NLA based on Alice's measurement results, respectively. \textbf{c} The EPR steerable regions parameterized by loss and gain with the NLA based on Alice's measurement results in a lossy channel. The blue and yellow regions are the two-way steerable region (I) and one-way steerable region (II), respectively.}
\label{figs4}
\end{figure*}

Here we show the theoretical results when the NLA based on Alice's measurement results is implemented in a lossy channel. As shown in Fig. 8a, the entanglement between Alice and Bob is enhanced after performing the NLA based on Alice's measurement results. For the distillation of EPR steering, both the steerablities of $\mathcal{G}^{A\rightarrow B}$ and $\mathcal{G}^{B\rightarrow A}$ are increased by the distillation, but the steerable regions are not changed which is different from the results with the NLA based Bob's measurement results [Figure 2b and 2c in the main text]. This is because the NLA based on Alice's measurement results can extent the steerable region of $\mathcal{G}^{A\rightarrow B}$ (as shown in the results of the NLA based on Alice's measurement results in a noisy channel [Figure 3d in the main text]), while the steerability of $\mathcal{G}^{A\rightarrow B}$ is robust against loss in a lossy channel.

\section{\label{sec:appendix5}The theoretical calculation of secret key rate for 1sDI QKD}

For a pure two-mode squeezed state, the variance of the amplitude and phase quadratures can be expressed by $V_{X}=V_{P}=(e^{2r}+e^{-2r})/2=V$. When Alice performs homodyne detection and Bob performs heterodyne detection, there will be a 50$\%$ loss introduced by a $50:50$ beam splitter in the heterodyne detection. So we have $V_{X_{A}}=V_{P_{A}}=V$,$%
V_{X_{B}}=V_{P_{B}}=VT+1-T$, $X_{A}X_{B}=-P_{A}P_{B}=\sqrt{T}%
(e^{-2r}-e^{2r})/2$, where $T=0.5$.

In the case of reverse reconciliation, the secret key rate for this 1sDI QKD protocol is bounded by Eq. (4) in the main text, i.e., $K^{\blacktriangleleft }\geq $log$_{2}{\frac{2}{e\sqrt{%
V_{P_{B}|P_{A}}V_{X_{B}|X_{A}}}}}$, in which  $V_{X_{B}\mid X_{A}}=V_{X_{B}}-\frac{\langle X_{A}X_{B}\rangle ^{2}%
}{V_{X_{A}}}=V_{P_{B}\mid P_{A}}$. And  $K^{\blacktriangleleft }$ is positive only for \bigskip $V_{X_{B}\mid X_{A}}V_{P_{B}\mid P_{A}}\leqslant (\frac{2}{e}%
)^{2}\approx 0.55$, which implies a squeezing parameter of $r\geqslant0.692$ (about $-6$ dB squeezing).

\end{document}